\documentclass[a4paper,fleqn,usenatbib]{mnras}

\usepackage{mathptmx}

\usepackage[T1]{fontenc}
\usepackage{ae,aecompl}

\usepackage{graphicx}	
\usepackage{amsmath}	
\usepackage{amssymb}	
\usepackage{times}
\usepackage{graphics, subfigure}
\usepackage{epsfig}
\usepackage{multirow}
\usepackage{ulem}
\usepackage{pdfpages}

\def\simlt{\mathrel{\hbox{\rlap{\hbox{\lower4pt\hbox{$\sim$}}}\hbox{$<$}}}}
\def\simgt{\mathrel{\hbox{\rlap{\hbox{\lower4pt\hbox{$\sim$}}}\hbox{$>$}}}}

\newcommand{\BV}{Brunt-V\"{a}is\"{a}l\"{a}\,}
\newcommand{\nc}{\bar{n}_{c}}
\newcommand{\Phiext}{\Phi_{\textrm{ext}}}

\newcommand{\beq}{\begin{equation}}
\newcommand{\eeq}{\end{equation}}
\newcommand{\bea}{\begin{eqnarray}}
\newcommand{\eea}{\end{eqnarray}}
\newcommand{\vv}{\vec v}

\newcommand{\vF}{\vec F}

\title[Dynamical tides reexpressed]{Dynamical Tides Reexpressed}

\author[D. Kushnir et al.]{
Doron Kushnir,$^{1,2}$\thanks{E-mail: kushnir@ias.edu}
Matias Zaldarriaga,$^{1}$
Juna A. Kollmeier$^{1,3}$
and Roni Waldman$^{4}$
\\
$^{1}$School of Natural Sciences, Institute for
Advanced Study, Princeton, New Jersey, 08540, USA\\
$^{2}$John N.\ Bahcall Fellow\\
$^{3}$ Observatories of the Carnegie Institution of Washington,
  813 Santa Barbara Street, Pasadena, CA 91101\\
$^{4}$Racah Institute of Physics, Hebrew University, Jerusalem, 9190401, Israel
}

\date{Accepted XXX. Received YYY; in original form ZZZ}

\pubyear{2017}

\begin{document}
\label{firstpage}
\pagerange{\pageref{firstpage}--\pageref{lastpage}}
\maketitle

\begin{abstract}
Zahn (1975) first put forward and calculated in detail the torque experienced by stars in a close binary systems due to dynamical tides. His widely used formula for stars with radiative envelopes and convective cores is expressed in terms of the stellar radius, even though the torque is actually being applied to the convective core at the core radius.  This results in a large prefactor, which is very sensitive to the global properties of the star, that multiplies the torque. This large factor is compensated by a very small multiplicative factor, $E_{2}$.  Although this is mathematically accurate, depending on the application this can lead to significant errors. The problem is even more severe, since the calculation of $E_{2}$ itself is non-trivial, and different authors have obtained inconsistent values of $E_{2}$. Moreover, many codes (e.g. BSE, StarTrack, MESA) interpolate (and sometimes extrapolate) a fit of $E_{2}$ values to the stellar mass, often in regimes where this is not sound practice. We express the torque in an alternate form, cast in terms of parameters at the envelope-core boundary and a dimensionless coefficient, $\beta_{2}$. Previous attempts to express the torque in such a form are either missing an important factor, which depends on the density profile of the star, or are not easy to implement. We show that $\beta_{2}$ is almost independent of the properties of the star and its value is approximately unity. Our formula for the torque is simple to implement and avoids the difficulties associated with the classic expression.
\end{abstract}


\section{Introduction}
\label{sec:Introduction}

The rate at which a binary system evolves to a state of a minimum kinetic energy (circular orbit, all spins aligned and synchronous with the orbital motion) depends on the dissipation of the tidal kinetic energy \citep[see][for a review]{Zahn2008}. When a radiative zone exists, radiative damping operates on the dynamical tide \citep{Zahn1975,Goodman1998}. It results from the excitation of internal gravity waves near the boundary of the convective zone and the radiative zone that then travel outwards (inwards) in the radiative envelope (core). 

In what follows, we concentrate on the synchronization of stars with radiative envelopes, as the derivations for circulation and for radiative cores is analogous. The torque applied to the star in this case is given by \citep{Zahn1975}
\begin{equation}
\label{eq:torque}
\tau=\frac{3}{2}\frac{G(qM)^2}{R}\left(\frac{R}{d}\right)^{6}E_{2}s^{8/3},
\end{equation}
where $M$ is the mass of the star, $qM$ is the mass of the companion, $R$ is the radius of the star, $d$ is the distance between the stars, $E_{2}$ is a parameter to be discussed below, $s=2|\omega-\Omega|\sqrt{R^{3}/GM}$ is the normalized apparent frequency of the tide, $\omega$ is the orbital angular velocity and $\Omega$ is the rotational angular velocity of the star. We consider the case $s^{-1}\gg1$, in which the forced oscillations in the envelope behave like a purely travelling wave, and the function $p(s)$ multiplying $\tau$ in \citet{Zahn1975} is $\approx1$. 

The torque is being applied to the convective core at radius $r_{c}$, but in order to cast the torque in terms of the global stellar properties (radius, $R$ and mass $qM$), \citet{Zahn1975} introduced the factor $E_2$ that depends strongly on $r_{c}/R$ and can thus compensate for this.  The numeric value of $E_2$ varies by many orders of magnitude for different stars owing to the strong dependence on the radius of the star, $\propto R^{9}$.

The physics of the internal gravity waves and how they lead to torques is now well understood. A simple order of magnitude of the effect can be found, for example, in \citet{Goldreich1989} \citep[see also][]{Savonije1984}:
\beq
\tau  \sim \rho \lambda \left(\omega-\Omega\right)^2 r^4 \left({\Phiext \over g r }\right)^2, 
\eeq
where $\Phi_{ext} = G (q M) r^2 /d^3$ is the perturbing tidal potential, $\rho$ is the density, $g$ is the gravitational acceleration, $\omega$ is the orbital angular velocity and $r$ is the radius. All quantities are evaluated at the boundary of the convective core and $\lambda$ is a length scale constructed using the derivative of the \BV frequency with radius at that location and determines the scale of variation of the waves in the radial direction. We show below that this estimate \citep[as well as][]{Savonije1984} lacks a factor $(1-\rho/\bar{\rho})^{2}$, where $\bar{\rho}$ is the average density of the core. This factor tends to decreases the torque for cores with density distribution which is close to uniform, and cannot be neglected. For example, for $n=3$ polytrope with $r_c/R$=0.1 this factor is $\approx7.9\cdot10^{-3}$, leading to an error of more than two orders of magnitude.

\citet{Goodman1998} repeated a simplified version of the \citet{Zahn1975} analysis for low mass stars where the core is radiative while the envelope is convective. These cases only differ by whether the waves propagate towards the surface of the star or towards the centre. As long as the waves get damped before they can get reflected back to the radiative-convective boundary there is no practical difference between these cases. However, it is not apparent that the torque derived by \citet{Goodman1998} is indeed identical to the expression derived by \citet{Zahn1975}. In Appendix~\ref{sec:torquescaling} we summarise the physics of the internal gravity waves and directly compare the formulas in \citet{Zahn1975}, \citet{Goldreich1989} and \citet{Goodman1998}. 
 
Although it is well understood that the torque does not depend on the radius of the star but on the location and properties at the boundary of the convective core, Equation~\eqref{eq:torque} is being routinely used in the literature. To reiterate the problem with doing so, a large factor of $(R/r_c)^9(M/M_c)^{-4/3} \sim10^{5}-10^{10}$, which is very sensitive to the the properties of the star, multiplies the torque by using the values at the stellar radius instead of at the core radius, and then a small factor of $10^{-5}-10^{-10}$ from $E_{2}$ is again multiplied to compensate for this large factor. While technically correct, in a given application this can lead to significant errors as both the large and the small factors are very sensitive to the properties of the stars and are not known exactly for many astrophysical systems of interest.  The problem is worse in practice, and different authors have obtained apparently inconsistent values of $E_{2}$ over time \citep{Zahn1975,Claret1997,Siess2013}. Moreover, \citet{Hurley2002} fit the $E_{2}$ values of \citet{Zahn1975} as a function of $M$, and this fit is adopted in many codes [e.g., BSE \citep{Hurley2002}, StarTrack \citep{Belczynski2008}, MESA \citep{Paxton2015}] to interpolate (and sometimes extrapolate) $E_{2}$ values, in a regime where this is wholly inappropriate. \citet{Siess2013} recognised this problem and pointed out that this procedure leads to at least an order of magnitude error for the value of the torque.  We emphasise here that the basic problem is the mere use of Equation~\eqref{eq:torque} and $E_{2}$, and we suggest an alternate expression for the torque, which is normalised to the core boundary with a dimensionless coefficient of order one. Previous attempts to express the torque in such a form are either missing the $(1-\rho/\bar{\rho})^{2}$ factor \citep{Savonije1984,Goldreich1989} or not easy to implement \citep{Goodman1998}.

In Section~\ref{sec:E2}, we review the calculation of the parameter $E_{2}$, and we resolve the apparent discrepancy between the inconsistent $E_{2}$ values in the literature, by showing that the value of $E_{2}$ can be accurately derived from a polytrope model. We further supplement $E_{2}$ values for main sequence (MS) models and Wolf--Rayet (WR) models calculated by the MESA code \citep[][see Appendix~\ref{sec:stellar models} for details]{MESA}. We use these WR models in a companion paper that analyses the consequences of the torque applied by a black hole to a WR star to the gravitational-wave emission from the subsequent merger of two black holes \citep{Kushnir:2016}. We demonstrate the sensitivity of $E_{2}$ on the properties of the star. In Section~\ref{sec:torque}, we express the torque with an alternate form, Equation~\eqref{eq:natural}, which is the main result of this paper. Readers not interested in the form of the torque that contains $E_{2}$, may skip directly to Section~\ref{sec:torque}.
 
\section{The value of $E_{2}$}\label{sec:E2}

The parameter $E_{2}$ introduced in \citet{Zahn1975} can be written as
\begin{eqnarray}\label{eq:E2 def}
E_{2}&=&\frac{3}{4\pi}\left(\frac{3}{2}\right)^{4/3}\frac{\Gamma^{2}(4/3)}{5}\left[\frac{\rho}{\bar{\rho}}\left(\frac{r}{g}\frac{dN^{2}}{d\ln r}\right)^{-1/3}\right]_{c}\times\nonumber\\
&&\left(\frac{M_{c}}{M}\right)^{2/3}\left(\frac{r_{c}}{R}\right)^{-3}H_{2}^{2}\nonumber\\
&\equiv& \bar{E}_{2}\left(\frac{r}{g}\frac{dN^{2}}{d\ln r}\right)^{-1/3}_c\equiv \bar{E}_{2}\nc,
\end{eqnarray}
where $\Gamma$ is the gamma function, the subscript $c$ indicates values at the convective core boundary,  $\bar{\rho}$ is the mean density inside the sphere of radius $r$, $N$ is the \BV frequency, and we assumed that $N^{2}_{c}=0$. The parameter $H_{2}$ is given by
\begin{equation}\label{eq:H2 def}
H_{2}=\frac{5}{[Y'(1)-Y(1)]X(x_{c})}\int_{0}^{x_{c}}\left(Y''-\frac{6Y}{x^{2}}\right)Xdx,
\end{equation}
where $Y$ is the solution of
\begin{equation}\label{eq:Y eq}
Y''-6\left(1-\frac{\rho}{\bar{\rho}}\right)\frac{Y'}{x}-6\left[1-2\left(1-\frac{\rho}{\bar{\rho}}\right)\right]\frac{Y}{x^{2}}=0,
\end{equation}
regular at the centre, $x$ is the normalised radius $r/R$, prime indicates a derivative with respect to $x$, and $X$ is the solution of 
\begin{equation}\label{eq:X eq}
X''-\frac{\rho'}{\rho}X'-6\frac{X}{x^{2}}=0,
\end{equation}
regular at the centre. Note that the factor $5/[Y'(1)-Y(1)]$ in Equation~\eqref{eq:H2 def} is sometimes replaced with $1/Y(1)$ \citep{Claret1997,Siess2013}, which is justified when the apsidal motion constant,
\begin{equation}\label{eq:k value}
k_{2}=\frac{6Y(1)-Y'(1)}{2[Y'(1)-Y(1)]},
\end{equation}
satisfies $k_{2}\ll1$. 

\begin{figure}
\includegraphics[width=0.45\textwidth]{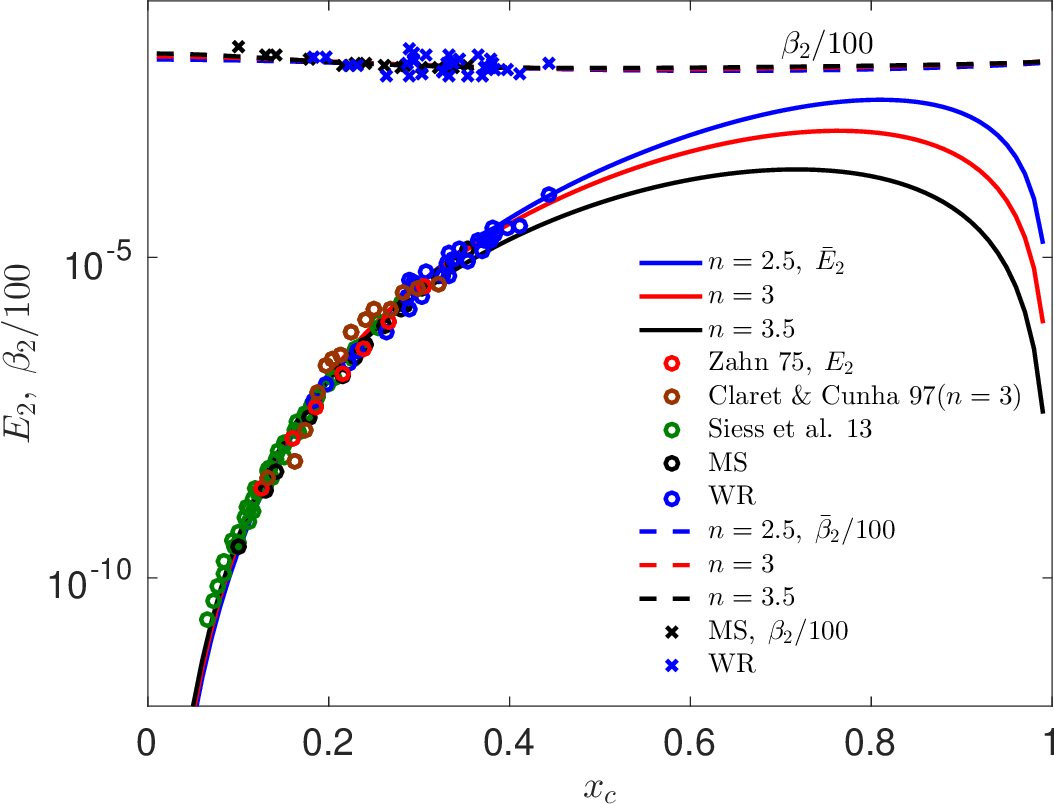}
\caption{The values of $E_{2}$ and $\beta_{2}/100$ as function of the normalised location of the convective core boundary $x_c=r_c/R$. Lines: Polytrope models, for which only $\bar{E}_{2}$ (solid) and $\bar{\beta}_2$ (dashed) can be calculated ($E_{2}$ and $\beta_2$ up to a factor $\nc=(r/g\cdot dN^{2}/d\ln r)^{-1/3}$, respectively); with blue, red and black for $n=2.5,3$ and $3.5$, respectively. Circles: $E_{2}$ values for different models; red:  \citet{Zahn1975} for the unpublished, simplified $1.6-15\,M_{\odot}$ MS stellar models of Aizenman, brown: \citet{Claret1997} for the MS stellar models of \citet{Claret1995,Claret1995b} over the mass range of $1.6-32\,M_{\odot}$, where the $x_{c}$ values are estimated by assuming $n=3$ polytrope, green: \citet{Siess2013} for the MS stellar models of \citet{Siess2006} over the mass range of $2-20\,M_{\odot}$, black: our MS models, blue: our WR models. x's: $\beta_{2}/100$ for our models; black: MS, blue: WR. Our WR results are an extension of the previous MS results to larger $x_{c}$ values. The variation of $E_{2}$ over many orders of magnitude is a direct consequence of the choice of parameters in Equation~\eqref{eq:torque}. $\beta_{2}$ is the dimensionless coefficient in our alternate form for the torque, expressed by using the values at the core boundary (Equation~\eqref{eq:natural}). $\beta_2$ is almost independent of the properties of the star and its value is approximately unity.
\label{fig:E2}}
\end{figure}

The value of $E_{2}$ is completely determined by the density profile, $\rho(r)$, the location of the convective core boundary and $\nc$. Assuming that $\rho(r)$ can be reasonably approximated by a polytrope and that $\nc$ is of the order of unity (both assumptions hold for our MS and WR models), then it is useful to inspect $E_{2}$ as a function of $x_{c}$ for a given stellar model in comparison with the results from the relevant polytrope. In Figure~\ref{fig:E2} we show $E_{2}$ as a function of $x_{c}$ for the following:
\begin{enumerate}
\item The results of \citet{Zahn1975} for the unpublished, simplified $1.6-15\,M_{\odot}$ MS stellar models of Aizenman. 
\item The results of \citet{Claret1997} for the MS stellar models of \citet{Claret1995,Claret1995b} over the mass range of  $1.6-32\,M_{\odot}$. Since the results of \citet{Claret1997} are given as a function of $M_{c}/M$, we assumed $n=3$ polytrope to estimate the $x_{c}$ values.
\item The results of \citet{Siess2013} for the MS stellar models of \citet{Siess2006} over the mass range $2-20\,M_{\odot}$. 
\item Our results for the MS and WR models (see Appendix~\ref{sec:stellar models}). Some details regarding the numerical calculation of $E_{2}$ are given in Appendix~\ref{sec:XY}.
\item Polytrope models, for which only $\bar{E}_{2}$ can be calculated ($E_{2}$ up to a factor $\nc$).
\end{enumerate}

\begin{figure}
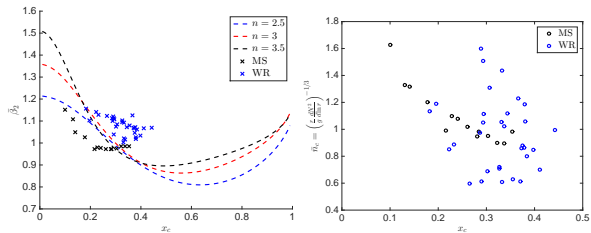

\includegraphics[width=0.45\linewidth]{beta2.eps}
\includegraphics[width=0.45\linewidth]{nc.eps}
\caption{Left-hand panel: $\bar{\beta}_{2}$ as function of the normalised location of the convective core boundary $x_c=r_c/R$. Lines: Polytrope models,  with blue, red, and black for $n=2.5,3$ and $3.5$, respectively. x's: our models; black: MS, blue: WR. The deviation between our MS and WR models and between the polytrope models are less than $20\%$, which is due to the small deviation of the stellar density profiles from polytropes. Right-hand panel:  $\nc=(r/g\cdot dN^{2}/d\ln r)^{-1/3}$ for our MS and WR models. The values of $\nc$ deviate from unity by less than a factor of $2$.}
\label{fig:beta2}
\end{figure}

We see that our MS results are consistent with the results of \citet{Siess2013}, of \citet{Claret1997}, and of \citet{Zahn1975}. For our MS and WR results, we can directly verify that the profiles are well described by polytropes with $n\approx2.5-3.5$, $\nc$ is in the range $\approx0.6-1.6$ (see right-hand panel of Figure~\ref{fig:beta2}) and that the $E_{2}$ estimate from the relevant polytrope is accurate to better than a factor $2$. Our WR results are an extension of the previous MS results to larger $x_{c}$ values. 

Most crucially, the variation of $E_{2}$ over many orders of magnitude is a direct consequence of the choice of parameters in Equation~\eqref{eq:torque}. In the next section, we rewrite Equation~\eqref{eq:torque} with physical parameters that minimise this large variance. 

\section{Alternate form for the torque}\label{sec:torque}

In this Section, we rewrite Equation~\eqref{eq:torque} to a form somewhat more natural from a theoretical point of view. The torque applied to the star is being applied to the convective core, so one would expect that the relevant quantity is the core radius $r_{c}=x_{c}R$ rather than the stellar radius $R$. As described in Appendix~\ref{sec:torquescaling}, Equation~\eqref{eq:torque} can be written as:
\begin{eqnarray}\label{eq:natural}
\tau &=& \frac{G(qM)^{2}}{r_{c}}\left(\frac{r_{c}}{d}\right)^{6}s_{c}^{8/3}\left[\frac{r_{c}}{g_{c}}\left(\frac{dN^{2}}{d\ln r}\right)_{r_{c}}\right]^{-1/3}\times\nonumber\\
&&\frac{\rho_{c}}{\bar{\rho}_{c}}\left(1-\frac{\rho_{c}}{\bar{\rho}_{c}}\right)^{2}\left[\frac{3}{2}\frac{3^{2/3}\Gamma^{2}(1/3)}{5\cdot6^{4/3}}\frac{3}{4\pi}\alpha_{2}^{2}\right],\nonumber\\
&\equiv&\beta_{2}\frac{G(qM)^{2}}{r_{c}}\left(\frac{r_{c}}{d}\right)^{6}s_{c}^{8/3}\frac{\rho_{c}}{\bar{\rho}_{c}}\left(1-\frac{\rho_{c}}{\bar{\rho}_{c}}\right)^{2}
\end{eqnarray}
where $s_{c}\equiv2|\omega-\Omega|\sqrt{r_{c}^{3}/GM_{c}}$, and $\alpha_2$ is an order-unity constant of proportionality that relates the displacement at the convective-radiative boundary in the dynamical tide relative to the equilibrium tide times the deviation from constant density. Note that all quantities in this expression are evaluated at the core boundary. Direct inspection of Equation~\eqref{eq:torque} shows that
\beq\label{eq:alpha res}
\alpha_{2}=\left(\frac{r_{c}}{R}\right)^{-5}\left(\frac{M_{c}}{M}\right)\left(1-\frac{\rho}{\bar{\rho}}\right)^{-1}H_{2}.
\eeq

In Figure~\ref{fig:E2}, we plot $\beta_{2}$ for our MS and WR models and for the polytrope models (for which one can calculate only $\bar{\beta}_{2}=\beta_{2}/\nc$ with $\nc=(r/g\cdot dN^{2}/d\ln r)^{-1/3}$). The value of $\beta_{2}$ is almost independent of $x_{c}$ and is approximately unity. Since the torque depends on high powers of the core radius and of the distance between the stars, one can take $\beta_{2}=1$ for most applications of this theory. We nevertheless plot, for completeness, the values of $\bar{\beta}_{2}$ in the left-hand panel of Figure~\ref{fig:beta2} using a linear scale. The deviation between our MS and WR models and between the polytrope models are less than $20\%$, which is due to the small deviation of the stellar density profiles from polytropes. The values of $\nc$ for our MS and WR models are shown in the right-hand panel of Figure~\ref{fig:beta2}, and they deviate from unity by less than a factor of $2$. 

In summary, we provide a simple alternative version of the \citet{Zahn1975} expression for the dynamical tidal torque experienced by a star in a close binary system.  Our alternate formulation references the core radius rather than the stellar radius and thus eliminates the need to compute $E_{2}$, and the associated pitfalls in doing so, for the majority of modern applications. Previous attempts to express the torque in such a form are either missing an important factor, which depends on the density profile of the star, or not easy to implement.

\section*{Acknowledgements}
We thank Ben Bar-Or, Jeremy Goodman, Boaz Katz, Roman Rafikov, and Eli Waxman for discussions. M.Z. is
supported in part by the NSF grants PHY-1213563, AST-
1409709 and PHY-1521097.   JAK gratefully acknowledges support from the Institute for Advanced Study.


\begin{appendix}

\section{Internal gravity waves}
\label{sec:torquescaling}

Here we aim to clarify the various formulae in the literature for the energy carried by the internal gravity waves excited at the boundary between the convective core and the radiative envelope. This will motivate using the location and other parameters of this boundary to express the torque. 

As a simple easily understood example, it is useful to consider the case of a plane parallel atmosphere where $z$ is the vertical direction and $g$ the acceleration of gravity. We are interested in internal gravity waves as those can have frequencies well below the dynamical frequency of the star that characterises the frequency of the sound waves. The restoring force for these waves is buoyancy and is characterised by the \BV frequency $N$:
\beq
N^2= g {d \over dz} \left({\ln p\over \Gamma_{1}} - \ln \rho\right),
\eeq 
where $p$ is the pressure and $\Gamma_{1}=(d\ln p/d\ln\rho)_{s}$. When a fluid element moves a distance $\delta z$ the resulting acceleration is given by $- N^2 \delta z \hat z$.

When considering internal gravity waves, one takes the density of the fluid as near constant with a value $\rho_c$. The acceleration of a fluid element is given by
\beq
{D \vv \over Dt} = - {1 \over \rho_c} \nabla p + b \hat z,
\eeq
where $b$ is the buoyancy and it satisfies:
\beq
{D b \over Dt} = - N^2 \hat z \cdot \vv.
\eeq
Finally, motions do not lead to compressions in the fluid as sound waves result in much higher frequency. Thus we also have
\beq
\nabla\cdot \vv = 0. 
\eeq
These equations are enough to recover the dynamics of internal gravity waves. One can combine them to obtain an equation for the $\hat z $ component of the velocity $w= \hat z \cdot \vv$, accurate to first order in the perturbation:
\beq
{\partial^2 \over \partial t^2 } \nabla^2 w = -N^2 \nabla^2_h w,
\eeq
where 
$\nabla^2_h = \nabla^2 - {\partial^2 / \partial z^2 }$ is the Laplacian in the horizontal direction. In the case where $N$ is constant one can easily derive the dispersion relation:
\beq
\omega^2 = N^2 {k_\perp^2 \over k^2},
\eeq
where $\omega$ is the angular frequency, $k_{\perp}^{2}=k_{x}^{2}+k_{y}^{2}$ and $k^{2}=k_{x}^{2}+k_{y}^{2}+k_{z}^{2}$. Note that the frequency decreases as $k_z$ increases, such that the group velocity is opposite in sign to the phase velocity. To have excitation of frequencies well below $N^2$ one needs ${k_\perp^2 \ll k^2}$; the modes need to have a very short wavelength in the vertical direction. For our application the $\hat z$ direction will be the radial direction of the star. The high spatial frequency of the oscillations will mean that tides will not easily excite these waves. As described in \citet{Goldreich1989}, the situation changes as one approaches the boundary of the radiative and convective regions as $N^2$ approaches zero there. There is a turning point for the waves at that location and the waves become evanescent in the convective zone. The waves vary more slowly in this part of the star. It is in there that tides can excite the waves. 

Because the crucial dynamics for the excitation of the waves happens near the radiative-convenctive boundary, which is near the turning point where $\omega^2 = N^2$, we cannot treat $N^2$ as constant. To get an analytic handle one can approximate $N^2$ as varying linearly with $z$. This was the approach adopted by \citet{Zahn1975} for massive stars and by \citet{Goodman1998} for low mass stars. At the level of discussion here, the main difference between the cases is whether the waves propagate towards the surface of the star or towards the centre. However, as long as the waves get damped before they can get reflected back to the radiative-convective boundary there is no practical difference between these cases for the present discussion. 

To analyse this case we approximate $N^2$ as:
\beq
N^2= N_0^2 + {d N_{0}^2 \over dz}  z.
\eeq
We will keep the $z$ dependence of all quantities explicitly and use a Fourier decomposition in the $x-y$ plane. We write for example:
\beq
w = W(z) e^{i (k_{x}x+k_{y}y - \omega t)}.
\eeq 
The equation for $w$ becomes:
\beq
W^{\prime\prime} + k_\perp^2 \left({N^2 \over \omega^2} -1\right) W = 0.
\eeq
The turning point is located were $N^2 = \omega^2$. In the case of a linear dependence of $N^2$ the equation becomes:
\beq\label{Weq}
W^{\prime\prime} + \tilde k^2 (z + z_0) W = 0, 
 \eeq
 where we introduced:
\bea
\tilde k^2 &=& { k_\perp^2 \over \omega^2} \left({d N^2\over  dz}\right)_{c} \nonumber \\
 z_0 &=& (N^2_0 - \omega^2) \left({d N_0^2\over  dz}\right)^{-1}.
\eea
The solutions of equation (\ref{Weq}) can be written in terms of Airy functions:
\bea
W(z) &=& a  \textrm{Ai}\left(-\frac{z + z_0}{\lambda}\right) + b  \textrm{Bi}\left(-\frac{z + z_0}{\lambda}\right) \nonumber \\
\lambda &=&  \tilde k^{-2/3} = \left({ k_\perp^2 \over \omega^2} {d N_0^2\over  dz}\right)^{-1/3}.
\eea
Assuming that waves excited at the radiative convective-boundary get damped before being reflected back, then $W$ describe outgoing waves, which are given by
\beq
W(z) = \tilde a\  \left[-i \textrm{Ai}\left(-\frac{z + z_0}{\lambda}\right) + \textrm{Bi}\left(-\frac{z + z_0}{\lambda}\right) \right],
\eeq
since $\textrm{Bi}(x)$ differs in phase from $\textrm{Ai}(x)$ by $\pi/2$ as $x\rightarrow\infty$ and the group velocity is opposite in sign to the phase velocity for these waves.

To obtain the normalisation $\tilde a$ one needs to compute how the equilibrium tide excites the waves at the radiative-convective boundary. The detailed calculation can be found in \citet{Zahn1975} for massive stars and in \citet{Goodman1998} or low mass stars, but the order of magnitude that will allow us to understand the dependence on the properties of the star can be found in \citet{Goldreich1989}.

To make this connection more explicit we can express the constant $\tilde a$ in terms of the displacement it implies at the turning point, $\delta z(-z_0) = {i \over \omega} W(-z_0)$. It is convenient to write the expression for the derivative of $\delta z$ 
\beq
\left.{ d \delta z \over dz }\right|_{-z_0} = {i \over \omega} W'(-z_0) =  -{i   \tilde a \over \lambda \omega} \  {2 \over 3^{1/3} \Gamma(1/3)}  {( 3^{1/2}+i) \over 2},
\eeq
or equivalently:
\beq
\tilde a =  { i \lambda \omega } \  { 3^{1/3} \Gamma(1/3)\over 2 } {( 3^{1/2} -i )\over 2} \left.{ d \delta z \over dz }\right|_{-z_0}.
\eeq
We have used the fact that $\textrm{Ai}^\prime(0) = - 1/3^{1/3} \Gamma(1/3)$ and  $\textrm{Bi}^\prime(0) =  3^{1/6}/ \Gamma(1/3)$.

Once we have the form of the outgoing waves we need to calculate the energy flux. The energy density is the sum of kinetic energy and potential energy that describes the buoyancy force. The energy conservation equation to first order in the perturbation is:
\bea
{\partial u \over \partial t} + \nabla \cdot \vF &=& 0 \nonumber \\
u &=&  {1\over 2 } \rho_0 (\vv^2 + {b^2 \over N^2}) \nonumber \\
\vF &=& \vv p,
\eea  
where $u$ is the energy density and $\vF$ the energy flux.  It is straightforward to use the linearized equations for the waves to show that this conservation law is satisfied. Thus to compute the energy flux carried by the waves away from the boundary (in the $\hat z$ direction) we need to calculate the time average of $w p$.  

For the pressure we write:
\beq
p = P(z) e^{i (k_{x}x+k_{y}y - \omega t)}.
\eeq 
Substituting this form into the equations of motion we find:
\beq
P = {i \omega \rho_c \over k_\perp^2} W^\prime, 
\eeq
thus the energy flux per unit area is given by:
\beq
{d L \over d A} = {1 \over 2} {\rm Re}( P W^*) =   -{\omega \rho_c \over 2 k_\perp^2} {\rm Im}( W^\prime W^*).
\eeq
Using the identity $\textrm{Ai}(x)\textrm{Bi}'(x)-\textrm{Ai}'(x)\textrm{Bi}(x)=\pi^{-1}$ \citep{Abramowitz}, we get: 
\beq
{d L \over d A} =  {\omega \rho_c \over 2 k_\perp^2} {|\tilde a|^2 \over \pi \lambda} =  { 3^{2/3} \Gamma^{2}(1/3)\over 8 \pi }  {\lambda \omega^3  \rho_c \over k_\perp^2}   
\left( \left.{ d \delta z \over dz }\right|_{-z_0} \right)^2.  
\eeq

We transform now to the spherical case, assuming no rotation of the star. We identify the $z$ direction with the radial direction and use $k_\perp^2 = l(l+1)/r^2$. With this replacement the scale $\lambda$ is given by:
\beq
\lambda = \left( {l(l+1) \over \omega^2 } {dN^2 \over d\ln r}\right)^{-1/3} r.
\eeq
The amplitude of the wave was determined at the turning point, which now we identify with the radiative-core boundary, $r_{c}$. This is justified, since $dN^{2}/d\ln r$ is on the order of the dynamical frequency of the star, and we have $\lambda/r\ll1$. To get the energy flux we need to integrate over the area of the core, which introduces another factor  $\int d\Omega r_{c}^2  |Y_{lm}|^2 = r_{c}^2$. After this translation our expression for the luminosity agrees with Equation (13) of \cite{Goodman1998}, which reads:
\bea
\label{eq:Goodman}
L ^G &=&   { 3^{2/3} \Gamma^{2}(1/3)\over 8 \pi }  [l(l+1)]^{-4/3} \omega^{11/3} \nonumber\\
&\times& \left[\rho r^5 \left({dN^2 \over d\ln r}\right)^{-1/3} \left({ d \delta r \over dr }\right)^2  \right]_{r=r_c}.
\eea

The order of magnitude of the displacement at the radiative convective boundary is given by the value for the equilibrium tide, the ratio of the external tidal potential $\Phiext\propto -G (q M) r^2 / d^3$ to the local gravitational acceleration $g$, times the deviation from constant density (see exact derivation below):
\beq
\delta r \propto - {\Phi_{ext} \over g}\left(1-\frac{\rho}{\bar{\rho}}\right),
\eeq 
where $\bar{\rho}$ is the mean density inside the sphere of radius $r$, or equivalently
\beq
{d \delta r \over dr} = \alpha {\Phi_{ext} \over g r}\left(1-\frac{\rho}{\bar{\rho}}\right),
\eeq 
where $\alpha$ is a proportionality constant. To get this proportionality constant, which is of order unity, requires solving the forced equations for the modes \citep{Zahn1975,Goodman1998}. However it is clear from the expression for $L$ that it only depends on the properties of the star near the radiative-convective boundary. Thus if we use the location, density and enclosed mass of this region to make dimensionless the expression of the energy flux, all we are doing when we solve the details of the modes across the star is calibrating a numerical constant of order one. 

Equation~\eqref{eq:Goodman} is also equivalent to equation (19) in \citet{Goldreich1989}:
\beq
L^{G}\sim\rho\lambda\omega^{3}r^{2}\left(\Phiext \over g\right),
\eeq
although without the factor $(1-\rho/\bar{\rho})^2$. Similar equation was derived by \citet{Savonije1984}, once again without the $(1-\rho/\bar{\rho})^2$ factor. This factor tends to decreases the torque for cores with density distribution which is close to uniform, and cannot be neglected. For example, for $n=3$ polytrope with $r_c/R$=0.1 this factor is $\approx7.9\cdot10^{-3}$, leading to an error of more than two orders of magnitude.

Next we make the connection with the work of \citet{Zahn1975}, where the radiated energy is given there in section 2.e. The flux of energy far away from the radiative convective boundary is computed using the asymptotic from of the modes far from the boundary, but as explained in \citet{Zahn1975} the waves are excited interior to this region. Solutions for the excited waves are found by matching solutions in the various regions of the star. In our notation equation (2.50) of \citet{Zahn1975} reads
\beq
L^Z = {1\over 2} {\omega ^2 R^3 K_0^2 \over [l(l+1)]^{1/2}},
\eeq
where $R$ is the stellar radius and because we want to make the connection with the notation of \citet{Goodman1998} we have used spherical harmonics rather than Legendre polynomials to define the modes and thus we have removed a factor of $4 \pi (l+m)! / (2 l+1) (l-m)!$. We have also assumed that in equation (2.50) of \citet{Zahn1975} $\gamma=1$, valid in the regime where waves do not reflect back to the radiative-convective boundary. The constant $K_0$ is given by equation (2.40) of \citet{Zahn1975}:   
\bea\label{K0}
K_0 &=& {\sqrt{3} \Gamma(4/3) \over 2 \sqrt{\pi}} \left( {v \over 3}\right)^{-5/6}\rho_c^{1/2}  \left[\left( {-g A \over x^2} \right)^\prime\right]^{1/4}_c \nonumber\\
&\times&\int_0^{x_c} \left[\left({x^2 \Psi \over g } \right)^{\prime\prime} - {l(l+1) \over x^2 } \left({x^2 \Psi \over g } \right)\right] {X \over X_c} dx,
\eea
where quantities with a subscript $c$ (rather than $f$)  are evaluated in the convective-radiative boundary, $x=r/R$ with $R$ the radius of the star,  primes are derivatives with respect to $x$,  $\Psi$ is the perturbation in the potential and $X$ is the solution of
\beq
X''-\frac{\rho'}{\rho}X'-l(l+1)\frac{X}{x^{2}}=0
\eeq 
regular at the centre. In terms of our notation, the parameter in $K_0$ are:
\bea
-g A &=& N^2 \nonumber \\
v^2 &=& {l(l+1) \over r_{c}^3 \omega^2} \left({d N^2 \over d\ln r} \right)_{r=r_{c}} R^3 =  \lambda^3 R^3 \nonumber \\
\left( {-g A \over x^2} \right)_c^\prime &=& {\omega ^2 \over l(l+1)} v^2. 
\eea
In order to estimate the integral in equation (\ref{K0}) we use the proportionality of $x^{2}\Psi/g$ to the function $Y(x)$ defined by \citet{Zahn1975}, which is the solution of
\begin{equation}
Y''-6\left(1-\frac{\rho}{\bar{\rho}}\right)\frac{Y'}{x}-\left[l(l+1)-12\left(1-\frac{\rho}{\bar{\rho}}\right)\right]\frac{Y}{x^{2}}=0
\end{equation}
regular at the centre. The integral is therefore
\beq
\int_0^{x_c} 6\left(1-\frac{\rho}{\bar{\rho}}\right)\left[\frac{1}{x}\left({x^2 \Psi \over g } \right)^{\prime} - \frac{2}{x^{2}}\left({x^2 \Psi \over g } \right)\right] {X \over X_c} dx.
\eeq
Since the integral is dominated by $x\approx x_c$ and the perturbation in the potential is dominated by the external tide $\Phiext$, it is useful to define $\alpha$:
\bea\label{eq:def alpha}
\alpha &=&  \left({g \over \Phiext x_c}\right)_{x_c} \left(1-\frac{\rho}{\bar{\rho}}\right)^{-1}_{x_c} \nonumber\\
&\times& \int_0^{x_c} \left[\left({x^2 \Psi \over g } \right)^{\prime\prime} - {l(l+1) \over x^2 } \left({x^2 \Psi \over g } \right)\right] {X \over X_c} dx
\eea
where $\alpha$ is a dimensionless parameter of order one. Replacing this definition into $L^Z$ we get:  
\bea
L ^Z &=&   { 3^{2/3} \Gamma^{2}(1/3)\over 8 \pi }  [l(l+1)]^{-4/3} \omega^{11/3}\alpha^2\nonumber\\
&\times& \left[\rho r^5 \left({dN^2 \over d\ln r}\right)^{-1/3} \left({\Phiext \over g r}\right)^2  \left(1-\frac{\rho}{\bar{\rho}}\right)^{2}\right]_{r=r_c}.
\eea 
Thus this is identical to the formula in \citet{Goodman1998} where $\alpha$ is the constant of proportionality that relates the displacement at the convective-radiative boundary in the dynamical tide relative to the equilibrium tide times the deviation from constant density. 
 
Finally in the main text we are interested in the torque carried away by the waves $\tau$, which is simply given by $\tau=2L/ \omega$. By specifying to $l=2$, such that $\Phiext=-(6\pi/5)^{1/2} G (q M) r^2 / d^3$, and replacing the angular frequency $\omega$ with the apparent angular frequency of the tide $2(\omega-\Omega)$, we get
\bea
\tau &=& \frac{G(qM)^{2}}{r_{c}}\left(\frac{r_{c}}{d}\right)^{6}s_{c}^{8/3}\left[\frac{r_{c}}{g_{c}}\left(\frac{dN^{2}}{d\ln r}\right)_{r_{c}}\right]^{-1/3}\frac{\rho_{c}}{\bar{\rho}_{c}}\left(1-\frac{\rho_{c}}{\bar{\rho}_{c}}\right)^{2}\nonumber\\
&\times&\left[\frac{3}{2}\frac{3^{2/3}\Gamma^{2}(1/3)}{5\cdot6^{4/3}}\frac{3}{4\pi}\alpha^{2}\right],
\eea
where $s_{c}\equiv2|\omega-\Omega|\sqrt{r_{c}^{3}/GM_{c}}$. The important point is that all quantities in this expression are evaluated at the core boundary. 
 
We will use this expression in the main text and to re-express the torque in the formulas of Zahn.  

\section{Stellar models}
\label{sec:stellar models}

We follow the evolution of the stellar models using the publicly available package MESA version 6596 \citep{MESA,2013ApJS..208....4P,Paxton2015}. For the MS models, we vary the zero--age main sequence (ZAMS) mass  between $[1.6,40] M_{\odot}$, thus covering the mass range of main--sequence stars with the convective core and the radiative envelope. All MS models had solar metallicity ($Z=0.02$) and no rotation. 

For the WR models, we aim at covering a wide range of masses during the WR phase $\approx[5,30] M_{\odot}$, and for that we select from a set of models having ZAMS masses in the range of $[40,100] M_{\odot}$, metallicities between $[0.01,1] Z_{\odot}$ and initial rotation between $[0.4,0.6]$ of breakup rate. We use the WR model profiles during the epoch where time to core-collapse is greater than $10^4\,\textrm{yr}$ and the stellar radius is smaller than $2 R_{\textrm{SM}}$, where $R_{\textrm{SM}}$ is the WR radius according to \citet{Schaerer1992}.

In all models, mass--loss was determined according to the `Dutch' recipe in MESA, combining the rates from \cite{2009A&A...497..255G, 1990A&A...231..134N, 2000A&A...360..227N, 2001A&A...369..574V}, with a coefficient $\eta=1$, convection according to the Ledoux criterion, with mixing length parameter $\alpha_{\textrm{mlt}}=2$, semi--convection efficiency parameter $\alpha_{\textrm{sc}}=0.1$ \citep[eq. 12]{2013ApJS..208....4P} and exponential overshoot parameter $f=0.008$ \citep[eq. 2]{MESA}. For the atmosphere boundary conditions we use the `simple' option of MESA \citep[eq. 3]{MESA}.

\section{Numerical integration of the functions $Y$ and $X$}
\label{sec:XY}

In the limit $x\rightarrow0$, we can write Equation~\eqref{eq:Y eq} as
\begin{equation}\label{eq:Y eq limit}
Y''-6\frac{Y}{x^{2}}=0,
\end{equation}
such that we can choose $Y(x)=x^{3}$ for the regular solution at the centre. Defining $Y\equiv f(x)x^{3}$, we get an equation for $f$
\begin{equation}\label{eq:f eq}
f''+6\frac{\rho}{\bar{\rho}}\frac{f'}{x}-6\left(1-\frac{\rho}{\bar{\rho}}\right)\frac{f}{x^{2}}=0,
\end{equation}
with $f(0)=1$. Expanding near $x=0$ as $f\approx1+a_{1}x+a_{2}x^{2}$ and $\rho\approx\rho_{0}+\rho_{0}'x+\rho_{0}''x^{2}/2$ we get
\begin{eqnarray}\label{eq:f limit}
\frac{\rho}{\bar{\rho}}&\approx&1+\frac{1}{4}\frac{\rho_{0}'}{\rho_{0}}x+\frac{1}{5}\frac{\rho_{0}''}{\rho_{0}}x^{2}-\frac{3}{4}\left(\frac{\rho_{0}'}{\rho_{0}}\right)^{2}x^{2}, \nonumber \\
&\Rightarrow&2a_{2}+6\left(1+\frac{1}{4}\frac{\rho_{0}'}{\rho_{0}}x\right)\left(\frac{a_{1}}{x}+2a_{2}\right)\nonumber\\
&+&\left[\frac{3}{2}\frac{\rho_{0}'}{\rho_{0}}+\frac{6}{5}\frac{\rho_{0}''}{\rho_{0}}x-\frac{9}{2}\left(\frac{\rho_{0}'}{\rho_{0}}\right)^{2}x\right]\left(\frac{1}{x}+a_{1}\right)=0.
\end{eqnarray}
Equating order by order we get
\begin{eqnarray}\label{eq:f limit 2}
f'(0)&=&-\frac{1}{4}\frac{\rho_{0}'}{\rho_{0}},\nonumber \\
f''(0)&=&\frac{3}{4}\left(\frac{\rho_{0}'}{\rho_{0}}\right)^{2}-\frac{6}{35}\frac{\rho_{0}''}{\rho_{0}}.
\end{eqnarray}
For numerical integration we can begin at some small $x_{\min}$ and estimate the boundary conditions using Equation~\eqref{eq:f limit 2}.

Similarly, in the limit $x\rightarrow0$, we can write Equation~\eqref{eq:X eq} as
\begin{equation}\label{eq:X eq limit}
X''-6\frac{X}{x^{2}}=0,
\end{equation}
such that we can choose $X(x)=x^{3}$ for the regular solution at the centre. Defining $X\equiv g(x)x^{3}$, we get an equation for $g$
\begin{equation}\label{eq:g eq}
g''+\left(\frac{6}{x}-\frac{\rho'}{\rho}\right)g'-3\frac{\rho'}{\rho}\frac{g}{x}=0,
\end{equation}
with $g(0)=1$. Expanding near $x=0$ as $g\approx1+b_{1}x+b_{2}x^{2}$ we get (assuming $\rho_{0}'\ne0$)
\begin{eqnarray}\label{eq:g limit}
\frac{\rho'}{\rho}&\approx&\frac{\rho_{0}'}{\rho_{0}}\left(1-\frac{\rho_{0}'}{\rho_{0}}x+\frac{\rho_{0}''}{\rho_{0}'}x\right), \nonumber \\
&\Rightarrow&2b_{2}+(b_{1}+2b_{2}x)\left(\frac{6}{x}-\frac{\rho_{0}'}{\rho_{0}}\right)\nonumber\\
&-&3\frac{\rho_{0}'}{\rho_{0}}\left(1-\frac{\rho_{0'}}{\rho_{0}}x+\frac{\rho_{0}''}{\rho_{0}'}x\right)\left(\frac{1}{x}+b_{1}\right)=0.
\end{eqnarray}
Equating order by order we get
\begin{eqnarray}\label{eq:g limit 2}
g'(0)&=&\frac{1}{2}\frac{\rho_{0}'}{\rho_{0}},\nonumber \\
g''(0)&=&-\frac{1}{7}\left(\frac{\rho_{0}'}{\rho_{0}}\right)^{2}+\frac{3}{7}\frac{\rho_{0}''}{\rho_{0}}.
\end{eqnarray}
One can verify that Equation~\eqref{eq:g limit 2} is also correct for the case $\rho_{0}'=0$. For numerical integration we can begin at some small $x_{\min}$ and estimate the boundary conditions using Equation~\eqref{eq:g limit 2}.

The derivation of the boundary conditions by \citet{Siess2013} is erroneous, since they did not expand $\rho$ near $x=0$, but rather used the value at $x=0$, which is inconsistent with the order of expansion. That leads to wrong values of $f''(0)$ and $g''(0)$, but has a negligible effect on the $E_{2}$ values.  

For a polytrope with index $n$, the dimensionless density of the star, $\theta$, satisfies the Lane-Emden equation of index $n$:
\begin{equation}\label{eq:Lane}
\frac{1}{\xi^{2}}\frac{d}{d\xi}\left(\xi^{2}\frac{d\theta}{d\xi}\right)=-\theta^{n},
\end{equation}
where $\xi$ is the radius and $\xi_{1}$ is the stellar radius ($\theta(\xi_{1})=0$). In the limit $\xi\rightarrow0$ we get $\theta\approx1-\xi^{2}/6$, leading to $\rho_{0}'=0$ and $\rho_{0}''/\rho_{0}=-n\xi_{1}^{2}/3$. 

\subsection{Analytical solutions for a special case}\label{sec:special}

For the case $\rho(x)=1-x^{2}$ we can find an analytical solutions for $f$ and $g$, which are useful for testing the numerical integration scheme. In this case we need to solve
\begin{equation}\label{eq:f special eq}
f''+\frac{30(1-x^{2})}{x(5-3x^{2})}f'-\frac{12}{5-3x^{2}}f=0,
\end{equation}
with $f(0)=1$ and $f'(0)=0$. The solution is
\begin{equation}\label{eq:f special}
f={}_{2}F_{1}\left(\frac{9-\sqrt{65}}{4},\frac{9+\sqrt{65}}{4};\frac{7}{2};\frac{3x^{2}}{5}\right),
\end{equation}
where ${}_{2}F_{1}$ is the hypergeometric function. 
The equation for $g$ is
\begin{equation}\label{eq:g special eq}
g''+\frac{6-4x^{2}}{x(1-x^{2})}g'+\frac{6}{1-x^{2}}g=0,
\end{equation}
with $g(0)=1$ and $g'(0)=0$. The solution is
\begin{equation}\label{eq:g special}
g={}_{2}F_{1}\left(\frac{3-\sqrt{33}}{4},\frac{3+\sqrt{33}}{4};\frac{7}{2};x^{2}\right).
\end{equation}

\end{appendix}

\bsp	
\label{lastpage}
\end{document}